\documentclass[runningheads]{llncs}
\usepackage[T1]{fontenc}
\usepackage{algorithm}
\usepackage{algpseudocode}  
\usepackage{graphicx}

\usepackage{mwe} 
\usepackage{csquotes}
\usepackage{float}
\usepackage{adjustbox}
\usepackage{multirow}
\usepackage{xcolor}
\usepackage{caption}
\usepackage{amsmath}
\usepackage{csquotes}

\begin{document}
\title{Clinically-Informed Preprocessing Improves Stroke Segmentation in Low-Resource Settings }
\titlerunning{Clinical Preprocessing for Stroke Segmentation in Low-Resource Settings}
%
\author{Juampablo E. {Heras Rivera}\inst{1}\orcidID{0000-0002-0205-6329} \and
Hitender Oswal\inst{1}\orcidID{0000-0002-4507-5466} \and
Tianyi Ren\inst{1}\orcidID{0000-0001-9548-6645} \and
Yutong Pan\inst{1} \and
William Henry\inst{1} \and
Caitlin M. Neher\inst{1}\orcidID{0009-0007-8249-6319} \and
Mehmet Kurt\inst{1}\orcidID{0000-0002-5618-0296}}
\authorrunning{Heras Rivera et al.}
%
\institute{University of Washington, Seattle WA 98105, USA \\
\email{\{jehr,hitender,tr1,ypan4,whenry1,neherc,mkurt\}@uw.edu}}

\maketitle              
\begin{abstract}
Stroke is among the top three causes of death worldwide, and accurate identification of ischemic stroke lesion boundaries from imaging is critical for diagnosis and treatment. The main imaging modalities used include magnetic resonance imaging (MRI), particularly diffusion weighted imaging (DWI), and computed tomography (CT)-based techniques such as non-contrast CT (NCCT), contrast-enhanced CT angiography (CTA), and CT perfusion (CTP). DWI is the gold standard for the identification of lesions but has limited applicability in low-resource settings due to prohibitive costs. CT-based imaging is currently the most practical imaging method in low-resource settings due to low costs and simplified logistics, but lacks the high specificity of MRI-based methods in monitoring ischemic insults. Supervised deep learning methods are the leading solution for automated ischemic stroke lesion segmentation and provide an opportunity to improve diagnostic quality in low-resource settings by incorporating insights from DWI when segmenting from CT. Here, we develop a series of models which use CT images taken upon arrival as inputs to predict follow-up lesion volumes annotated from DWI taken 2-9 days later. Furthermore, we implement clinically motivated preprocessing steps and show that the proposed pipeline results in a \textbf{38\%} improvement in Dice score over 10 folds compared to a nnU-Net model trained with the baseline preprocessing. Finally, we demonstrate that through additional preprocessing of CTA maps to extract vessel segmentations, we further improve our best model by \textbf{21\%} over 5 folds.

\end{abstract}

\section{Introduction}

Stroke is among the top three causes of death worldwide, with ischemic strokes accounting for over 87\% of cases \cite{strokefax}. In the acute (early) setting, medical imaging is critical for rapid accurate diagnosis, treatment triage, prognosis prediction, and secondary preventive precautions \cite{therole}. Diffusion weighted imaging (DWI) MRI and computed tomography (CT) are the primary imaging modalities used to understand stroke lesion progression and are typically used to obtain segmentations of the lesion boundaries. In the emergency setting, CT is always used due to its availability, low cost, and speed, while DWI is rarely available. 

DWI is approximately 4 to 5 times more sensitive in detecting acute stroke than non-contrast CT (NCCT), and is the gold standard as it can detect 95\% of hyperacute ischemic infarcts \cite{chalela2007mri, brazzelli2009mri, fiebach2002ctmri, gonzalez1999dwi}. DWI is capable of detecting acute brain infarction within 1 to 2 hours, while NCCT may be negative for the first 24 to 36 hours \cite{pantano19992436, therole}. Furthermore, automated segmentation of the ischemic core using deep learning methods with DWI as input is well-established, with Dice scores surpassing 80\% when compared to clinician-annotated labels \cite{isles22}. Despite strong evidence supporting DWI as superior to NCCT for confirming the diagnosis of acute stroke within the first 24 hours, logistical and financial issues limit its use in acute settings since most institutions find it challenging to reserve MRI scanners without delaying treatment. 

In low-resource settings, these issues are further exacerbated, as the availability of MRI in acute settings is incredibly limited. For instance, in Africa and many countries in South America, there is less than 1 MRI scanner per million people \cite{africa1M}. For context, in high-resource countries like the United States, there are over 40 MRI scanners per million people \cite{usa40M}. This means that although DWI is the most reliable imaging method for stroke lesion identification, it remains largely inaccessible in low-resource settings. As a result, CT remains the most prevalent imaging modality for acute ischemic stroke treatment around the world due to its simplified logistics and significantly lower operational costs. 

Although DWI remains widely unavailable in low-resource settings, supervised deep learning methods provide an opportunity to improve diagnostic quality in low-resource settings by incorporating insights from DWI data when segmenting stroke lesions from CT. Here, we present an approach for stroke lesion segmentation which uses CT scans from the acute ischemic stroke setting to predict follow-up stroke lesion volumes obtained from DWI 2-9 days after. For training and evaluation, we use the ISLES'24 \cite{isles24} challenge dataset, which provides longitudinal imaging data of stroke patients, including acute NCCT, CT angiography (CTA), CT perfusion (CTP), DWI, and ground truth hand-annotated lesion masks obtained from DWI. 

In comparison to MRI-based segmentation which requires minimal preprocessing, CT scans require significant preprocessing to remove irrelevant information and provide a clear learning signal for the segmentation model. Here, we develop a novel preprocessing pipeline to improve segmentation performance, considering the reasoning patterns used by clinicians. Although simple and computationally inexpensive, we show that the proposed pipeline results in a \textbf{45\%} improvement in performance when compared comparison to a model trained with baseline nnU-Net preprocessing. Finally, we demonstrate that through additional preprocessing of CTA maps to extract vessel segmentations, we can further improve our best model by \textbf{21\%} on average over 5-folds.

\section{Dataset}
The ISLES'24 challenge dataset was used for model training and evaluation. It consists of multi-center, multi-scanner imaging and tabular data from patients with large vessel occlusion (LVO) ischemic stroke. Imaging was acquired at two time points: at admission (acute) and at follow-up 2–9 days later (subacute).

Acute imaging includes the diagnostic CT trilogy: NCCT, CTA, and CTP. Additionally, it includes four CTP-derived maps: cerebral blood flow (CBF), cerebral blood volume (CBV), mean transit time (MTT), and time to maximum (TMax). These maps were generated using an FDA-approved clinical software (\texttt{icobrain CVA}), following motion correction and deconvolution processing. Follow-up imaging includes DWI and apparent diffusion coefficient (ADC) maps, which were used to derive stroke lesion core segmentations. Ground truth labels were generated using an ensemble segmentation pipeline from ISLES’22 \cite{isles22} and quality-checked by neuroradiologists.

Each imaging modality offers complementary information about stroke pathology. CTA uses contrast-enhancement to provide a view of cerebral vasculature, allowing detection of occlusions and collateral flow. CBF reflects the rate of blood delivery to brain tissue, CBV indicates the total blood volume within a region, MTT represents the average time it takes blood to pass through a voxel, and TMax captures delays in contrast arrival. These quantities are assessed together to infer tissue at risk of irreversible damage (penumbra), vs. the unrecoverable tissue (core). Follow-up DWI detects regions of restricted diffusion due to cytotoxic edema and is highly sensitive to infarcted tissue. Apparent Diffusion Coefficient (ADC) models how freely water molecules can move within a specific area of tissue and helps to confirm diffusion abnormalities. 

In total, the ISLES'24 dataset includes 250 sets of scans, with 150 sets for training (100 from the University Hospital of Munich and 50 from the University Hospital of Zurich), and 100 sets for testing (from undisclosed hospitals).

\section{Methods}
This section describes the components of our proposed method, which includes a clinically-guided preprocessing pipeline. This includes brain volume extraction with \texttt{SynthStrip} \cite{SynthStrip} (Section \ref{subsec:brainex}), clinically  motivated intensity windowing (Section \ref{subsec:intensity_windowing}), and segmentation of blood vessels from CTA (Section \ref{subsec:vesselseg}). Then, to obtain stroke lesion segmentations, a standard residual nnU-Net model \cite{nnunet} is used (Section \ref{subsec:nnunet}).  The models were trained with 5 preprocessed inputs, consisting of: CTA, MTT, TMax, CBV, and CBF maps, and one-channel binary ischemic core masks as output. For the best model, the CTA input was replaced with binary vessel segmentation maps obtained from CTA.


\subsection{Brain extraction}
\label{subsec:brainex}
The scans in the ISLES'24 brain imaging dataset contain non-brain structures such as the skull and background artifacts, which can hinder model training. To address this, we apply \texttt{SynthStrip}, a deep-learning-based brain-extraction tool trained on diverse synthetic images. \texttt{SynthStrip} allows for accurate and fast brain extraction of the entire dataset in minutes, a process that would have taken days for traditional brain extraction algorithms. First, we applied \texttt{SynthStrip} on the NCCT scans to obtain a brain masks. Then, we applied this brain mask to the other co-registered scans (CTA, and CTP-derived maps) to obtain skull-stripped versions of the data. 

\subsection{Vessel Segmentation using CTA}
\label{subsec:vesselseg}
CTA scans quickly and reliably add important information in cases of acute ischemic stroke. CTA shows the site of occlusion, the length of the occluded arterial segment, and the contrast-enhanced arteries beyond the occlusion as an estimate of collateral
blood flow \cite{knauth1997potential}. However, for deep learning models to effectively process CTA, it is beneficial to segment the vessel regions from CTA as binary masks and use these as inputs. Our proposed vessel segmentation pipeline is shown in Algorithm \ref{algo:vesselseg}.

\begin{algorithm}[H]
\label{algo:vesselseg}
\caption{Vessel Segmentation using CTA}
\begin{algorithmic}[1]

\Statex \textbf{Input:} $\mathbf{I}_{\text{CTA}}$: CTA volume, $\mathbf{I}_{\text{NCCT}}$: NCCT volume, $\mathbf{B}$: Brain mask from \texttt{SynthStrip}
\Statex \textbf{Params:} HU window $[0,400]$, thresholds $\tau_{\text{low}}{=}50$, $\tau_{\text{high}}{=}400$, min size $s_{\min}{=}25$
\Statex \textbf{Output:} $\mathbf{M}_{\text{vessel}}$: Binary vessel mask 
\Statex
\State $\mathbf{I}_{\text{CTA}} \gets \text{clip}(\mathbf{I}_{\text{CTA}}, 0, 400)$, $\mathbf{I}_{\text{NCCT}} \gets \text{clip}(\mathbf{I}_{\text{NCCT}}, 0, 400)$ \Comment{Clip intensities}
\State $\mathbf{I}_{\text{CTA}} \gets \mathbf{I}_{\text{CTA}} \odot \mathbf{B}$, $\mathbf{I}_{\text{NCCT}} \gets \mathbf{I}_{\text{NCCT}} \odot \mathbf{B}$ \Comment{Apply brain mask}
\State $\mathbf{D} \gets \mathbf{I}_{\text{CTA}} - \mathbf{I}_{\text{NCCT}}$ \Comment{Voxel-wise difference}
\State $\mathbf{D}(\mathbf{D} < \tau_{\text{low}}) \gets 0$, $\mathbf{D}(\mathbf{D} > \tau_{\text{high}}) \gets 0$ \Comment{Suppress low contrast and artefacts}
\State $\mathbf{M}_0 \gets (\mathbf{D} \ne 0)$ \Comment{Binary candidate mask}
\State $\{\mathcal{C}_k\} \gets \text{LabelConnectedComponents}(\mathbf{M}_0)$ \Comment{Connected components}
\State $\mathbf{M}_{\text{vessel}} \gets \bigcup_{k:\,|\mathcal{C}_k| \ge s_{\min}} \mathcal{C}_k$ \Comment{Keep components with size $\ge s_{\min}$ voxels}
\State \Return $\mathbf{M}_{\text{vessel}}$

\end{algorithmic}
\label{algo:vesselseg}
\end{algorithm}

\begin{figure}
    \centering
    \includegraphics[width=0.95\linewidth]{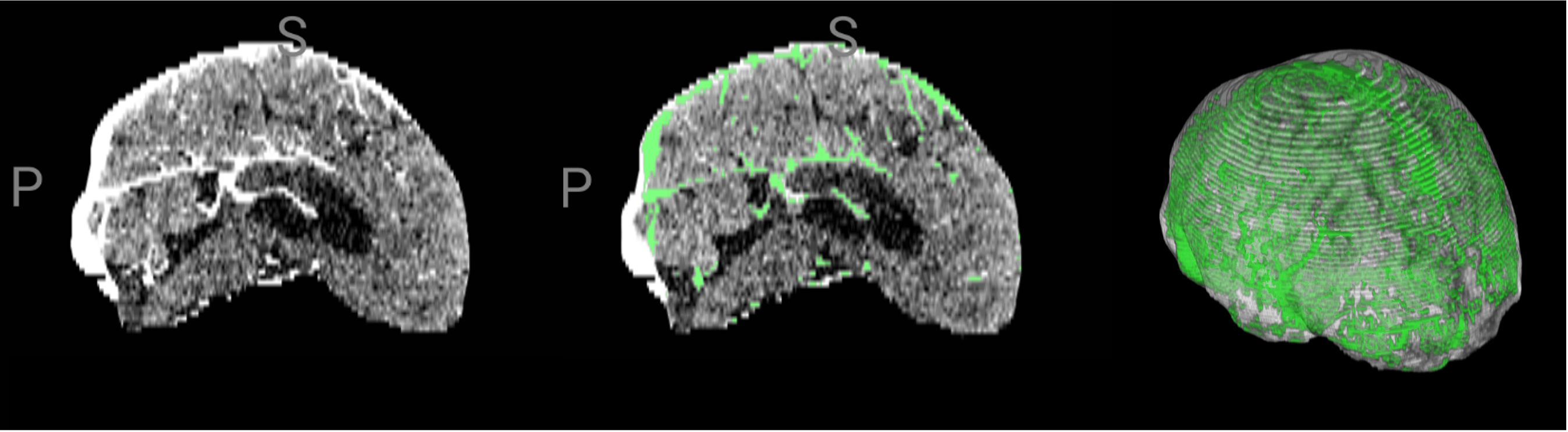}
    \caption{Sagittal view of a subject's CTA (left) and the same view with the vessel map extracted using Algorithm~\ref{algo:vesselseg} overlaid in green (middle). Right: 3D rendering of the CTA scan with the same vessel mask overlaid in green.}
    \label{fig:vessel}
\end{figure}

\textbf{Algorithm \ref{algo:vesselseg} description: }Starting from CTA and the co-registered NCCT, we first clip voxel intensities to $[0, 400]$ Hounsfield Units (HU) to supress extreme artifacts and noise. Then we apply a brain mask generated with \texttt{SynthStrip} to both scans to exclude structures outside of the brain. We then compute a voxel-wise difference between the CTA and NCCT, which highlights contrast-filled vessels while canceling parenchyma and remaining bone. Afterwards, we supress voxels with low ($<50$ HU) and high ($>400$ HU) contrast, to focus on the vessel regions. Then, we make a binary mask based on remaining nonzero voxels. Connected component analysis is performed on this mask, and only components containing at least 25 voxels are retained, producing the final binary vessel mask $\mathbf{M}_{\text{vessel}}$. An example of a CTA from a subject in the dataset and their overlaid vessel mask is shown in Figure \ref{fig:vessel}.

\subsection{Intensity Windowing}
\label{subsec:intensity_windowing}

\begin{figure}[H]
    \centering
    \includegraphics[width=0.75\linewidth]{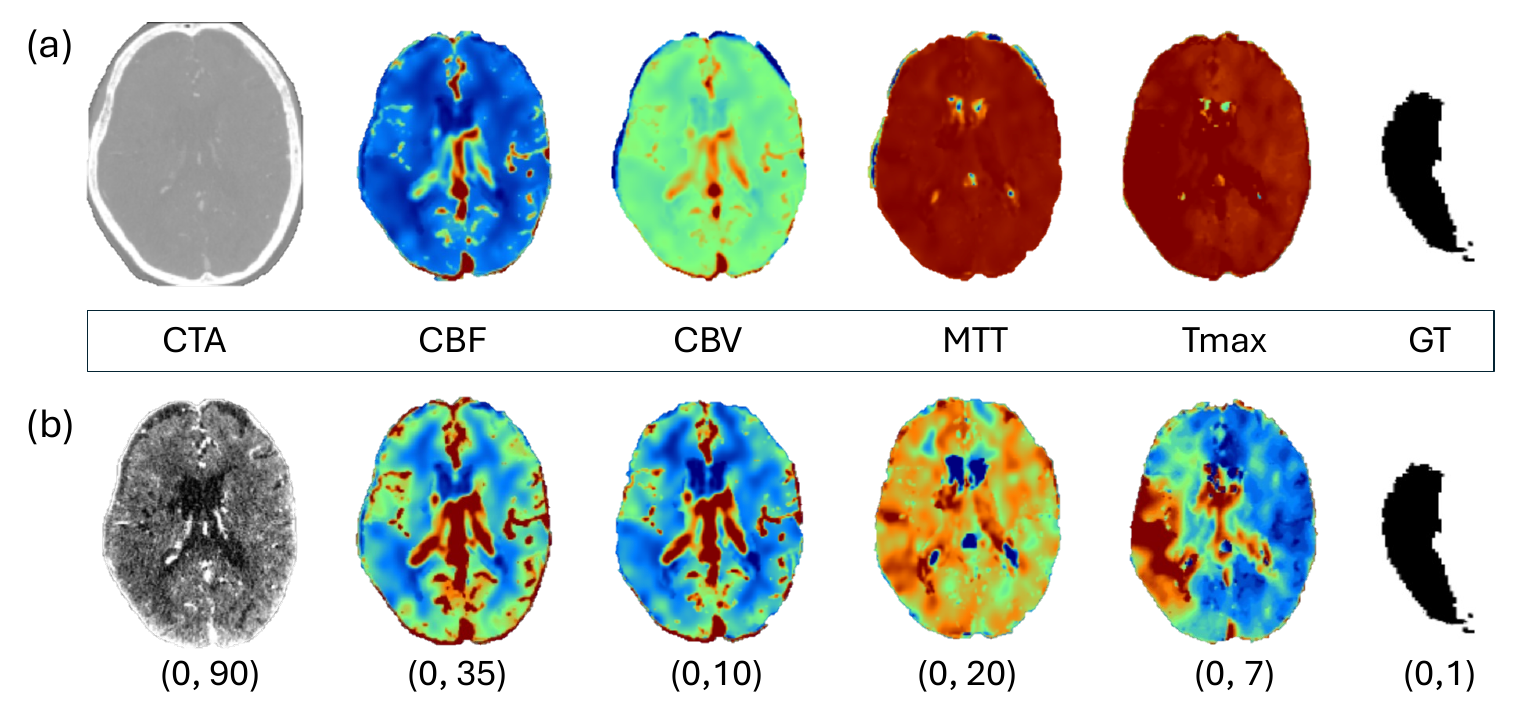}
    \caption{Imaging data from an example subject (a) before and (b) after preprocessing, with windowing bounds shown below (b). Over all images, the clinically-informed preprocessing increases visibility of the stroke lesion (GT).}
    \label{fig:pre_vs_post}
\end{figure}

To remove extraneous information from CT images, intensity value windowing was applied.  When clinical windowing guidelines were available in the literature, they informed the initial settings, and empirical adjustments were made to optimize model performance. The clinically established thresholds used for reference include: CBF~$<$~17~mL/100g/min~\cite{bandera2006cbfthreshold}; CBV~$>$~2~mL/100g~\cite{wintermark2006perfusion}; MTT~$>$~145\% of the contralateral baseline, where the 0--30~HU range captures the full variability~\cite{nukovic2023neuroimaging,alzahrani2023assessing,czap2021overview}; and Tmax~$>$~6~s~\cite{olivot2009optimal}. For CTA, the window was manually adjusted to enhance contrast between healthy and ischemic brain regions, as in~\cite{pulli2012acute}. For inputs with inconclusive clinical thresholds, such as CTA, MTT, and CBF, further widening of the selected windows beyond ranges found in the literature was performed to improve visibility of the ischemic lesion. The windowing bounds used for the proposed preprocessing strategy can be found in Table \ref{tab:ct_window_values} as \enquote{Clinical Window}, and an example of a subject's scans before and after windowing are shown in Figure \ref{fig:pre_vs_post}. After windowing, values were min-max normalized to the $[0,1]$ range, then 3-D histogram equalization was done on foreground voxels. Finally, all background voxels were assigned zero to remove artifacts outside of the brain.


\begin{table}[h]
\centering
\renewcommand{\arraystretch}{1.2}
\begin{adjustbox}{width=\linewidth}
\begin{tabular}{|c|c|c|c|}
\hline
\textbf{Modality} &
\textbf{Clinical Window} &
\textbf{nnU-Net CT Range} &
\textbf{\% of Range Kept} \\
\hline
CTA (HU)          & (0, 90)  & (\(-3.25\), 342.48)  & 26.0\,\% \\
CBF (mL/100g/min) & (0, 35)  & (1.42, 72.64)        & 49.2\,\% \\
CBV (mL/100g)     & (0, 10)  & (\(-10.31\), 19.35)  & 33.7\,\% \\
MTT (s)           & (0, 20)  & (\(-96.91\), 28.50)  & 15.9\,\% \\
Tmax (s)          & (0, 7)   & (\(-20.76\), 20.29)  & 17.1\,\% \\
\hline
\end{tabular}
\end{adjustbox}
\caption{Comparison of our clinically chosen intensity windows with the nnU-Net foreground intensity range (0.5$^{\text{th}}$–99.5$^{\text{th}}$ percentiles).  The last column shows what fraction of the nnU-Net range each clinical window retains.}
\label{tab:ct_window_values}
\end{table}


\subsection{nnU-Net}
\label{subsec:nnunet}
Despite the emergence of models like Transformers \cite{hatamizadeh2021swin} and diffusion frameworks \cite{ren2024re}, CNNs based on the U-Net architecture \cite{unet} remain state-of-the-art for medical image segmentation \cite{nnunet_sota}. The nnU-Net framework \cite{nnunet} automates hyperparameter tuning by adapting a standard U-Net to the training data, often outperforming manually tuned and novel models \cite{rivera2024ensemble}. For the models evaluated here, the 3D nnU-Net "ResEnc L” with a (56, 320, 256) patch size, Dice and cross-entropy loss, and the SGD optimizer (lr=0.01, momentum=0.99) was used.
For CT data, nnU-Net clips intensities to the $0.5-99.5$th percentiles of all foreground voxels (union of the label masks), then applies Z-score normalization using the mean and standard deviation of those same foreground intensities \cite{nnunet}. The clipped ranges kept by nnU-Net for the ISLES'24 dataset are in Table \ref{tab:ct_window_values} as \enquote{nnU-Net CT Range}.




\section{Results}


Two sets of experiments were conducted, one with 10-fold cross validation (Section \ref{subsec:tenfold}), and one with 5-fold cross validation (Section \ref{subsec:fivefold}). The model from the highest-performing fold of the 10-fold experiments was also submitted to the ISLES'24 challenge, where it achieved first place (Section \ref{subsec:isleswin}).
\subsection{10-fold CV experiments}
\label{subsec:tenfold}

Table 2 shows the results for the highest validation Dice score for the 10-fold cross validation experiment. On this fold, the standard nnU-Net preprocessing achieved a Dice score of 21.8\%. Applying custom windowing alone improved performance to 31.0\%, and combining custom windowing with Z-score normalization (as described in Section \ref{subsec:intensity_windowing}) further increased the validation Dice score to \textbf{31.8\%}.


\begin{table}[H]
\centering
{\normalsize
\renewcommand{\arraystretch}{1.2}
\begin{tabular}{|l|c|c|c|}
\hline
\textbf{Preprocessing Strategy} &
\textbf{Dice (mean ± SD)} &
\(\Delta\) (mean) &
\textbf{Dice (best fold)} \\
\hline
Baseline (nnU-Net default)                      & 0.162 (0.065) & --            & 0.218 \\
\textbf{+ Section \ref{subsec:intensity_windowing} Preproc.} & \textbf{0.224 (0.047)} & \(\mathbf{+38.4\%}\) & \textbf{0.318} \\
\hline
\end{tabular}
}
\caption{Dice scores for the baseline nnU-Net preprocessing and the clinically informed preprocessing (windowing + histogram equalization). Values are reported as the mean ± standard deviation over ten folds, together with the best-fold Dice score. }
\label{tab:preprocessing_dice}
\end{table}


\subsection{5-fold CV experiments}
\label{subsec:fivefold}

For these experiments, 5-fold cross-validation was used. The baseline model in Table~\ref{tab:vessel_vs_baseline_dice} applies the best preprocessing strategy from Section~\ref{subsec:tenfold}. In the improved setup, the windowed CTA input is replaced with binary vessel maps from Section~\ref{subsec:vesselseg}, resulting in a 21\% performance gain. Table~\ref{tab:vessel_vs_baseline_dice} reports results across all 5 folds for both approaches.

\begin{table}[H]
\centering
\renewcommand{\arraystretch}{1.15}
\setlength{\tabcolsep}{5pt}
\begin{tabular}{|l|c|c|c|c|c||c|c|}
\hline
\textbf{Dice Score} & \textbf{Fold 1} & \textbf{Fold 2} & \textbf{Fold 3} & \textbf{Fold 4} & \textbf{Fold 5} & \textbf{Mean} & \textbf{STD} \\
\hline
Best Preproc.                      & 0.2167 & \textcolor{blue}{0.3184} & 0.2181 & 0.1521 & 0.1696 & 0.2150 & 0.0647 \\
+ Vessel Segm.               & 0.2257 & \textcolor{blue}{0.3273} & 0.2129 & 0.2852 & 0.2495 & \textbf{0.2601} & \textbf{0.0466} \\
\hline
\(\Delta\) (\%)                    & +4.2 & +2.8 & -2.4 & +87.5 & +47.1 & \textbf{+21.0} & \textbf{-27.9} \\
\hline
\end{tabular}
\caption{Validation Dice scores for the best preprocessing pipeline from Section \ref{subsec:tenfold} versus the same model trained with the vessel segmentation replacing CTA, evaluated on the same 5-fold split.  \(\Delta\) gives the relative percentage change with respect to the best preprocessing pipeline from Section \ref{subsec:tenfold}. The best fold for both experiments is highlighted in \textcolor{blue}{blue}.}
\label{tab:vessel_vs_baseline_dice}
\end{table}

\subsection{ISLES'24 challenge submission}
\label{subsec:isleswin}
The model from the best fold as described in Section \ref{subsec:tenfold} was submitted to the ISLES'24 challenge \cite{isles24} and achieved first place \cite{wonisles}. Table \ref{table:test_metrics_leaderboard} shows test set evaluation metrics for the top 3 entries in the ISLES'24 leaderboard. 

\begin{table}[H]
\centering
\begin{adjustbox}{width=0.975\linewidth}
\begin{tabular}{|c|c|c|c|c|}
\hline
Team & Dice (\%) $\uparrow$ & AVD $\downarrow$ & F1 (\%) $\uparrow$ & ALCD $\downarrow$ \\
\hline
\textbf{Kurtlab (Ours)} & \textbf{28.50 (21.27)} & \textbf{21.23 (37.22)} & \textcolor{blue}{14.39 (21.19)} & \textcolor{blue}{7.18 (7.67)} \\
AMC-Axolotls & \textcolor{blue}{26.27 (24.73)} & \textcolor{blue}{21.31 (35.23)} & \textbf{14.94 (25.12)} & 7.66 (7.94) \\
Ninjas & 25.46 (19.08) & 26.29 (39.73) & 9.92 (13.46) & \textbf{5.98 (6.46)} \\
\hline
\end{tabular}
\end{adjustbox}
\caption{Test set evaluation metrics for the top 3 entries in the ISLES'24 leaderboard, reported as mean (standard deviation). Metrics shown are: Dice coefficient (Dice), Absolute volume difference 
 (AVD), F1-score, and Absolute lesion count difference (ALCD). Per column, \textbf{Bold} = best, \textcolor{blue}{blue} = second best. }
\label{table:test_metrics_leaderboard}
\end{table}

\section{Analysis}
The results of our experiments demonstrate that clinically informed preprocessing improves stroke lesion segmentation from CT imaging. In the first set of experiments, we show that applying skull-stripping and custom intensity windowing, followed by histogram equalization, improves model performance by $38.4\%$ compared to the baseline nnU-Net preprocessing. This is likely due to standard percentile-based preprocessing preserving irrelevant high-intensity regions (e.g., skull), while the stroke core typically occupies a narrow band within the overall intensity range. By using clinically informed preprocessing strategies, the model is able to focus on the anatomically and pathophysiologically relevant structures.  

In the second set of experiments, we observe that vessel structure is the most informative component of CTA for stroke lesion segmentation. By segmenting the vessels prior to training, we introduce a strong clinical prior, improving model performance. This improvement allows the model to reduce false negatives, especially in cases of LVOs. While our proposed preprocessing pipeline improves performance, it still exhibits high variance on the ISLES'24 test set (Table \ref{table:test_metrics_leaderboard}), suggesting robustness should be improved in future works. We expect this variance will  decrease as dataset size increases, as more training samples should improve generalization.

\section{Conclusions}
In low-resource settings, accurate prediction of ischemic core from CT scans is critical to reduce disparities in treatment quality in comparison to countries which have access to DWI. This paper presents a novel preprocessing strategy for stroke lesion segmentation, which improves the ability of models to predict subacute stroke lesions using CT imaging obtained in the acute phase. The presented analysis shows that standard preprocessing pipelines, such as those used in nnU-Net for CT scans, are insufficient to segment ischemic stroke lesions in CT. Here, we share the details of our preprocessing approach which increases the visibility of tissues of interest and allows the model to focus on clinically-relevant structures. The results of our preprocessing steps also allowed us to place first in the ISLES'24 challenge. Further studies in this direction will involve improving robustness of the segmentation approach by implementing more clinical priors in the preprocessing pipeline.

\begin{credits}
\subsubsection{\ackname} The work of Juampablo Heras Rivera was partially supported by the U.S. Department of Energy Computational Science Graduate Fellowship under Award Number DE-SC0024386.

\subsubsection{\discintname}
The authors have no competing interests to declare that are
relevant to the content of this article. 
\end{credits}
%
%
%
%




\newpage
\bibliographystyle{splncs04}
\bibliography{midl-samplebibliography}

\end{document}